\documentclass[twocolumn,preprintnumbers,amsmath,amssymb]{revtex4}
\usepackage{graphicx}
\usepackage{bm}
\newcommand{\rmf}{{\rm{f}}}
\newcommand{\bi}{\bibitem}
\newcommand{\be}[1]{\begin{equation} \label{#1} }
\newcommand{\bea}[1]{\begin{eqnarray} \label{#1} }

\newcommand{\bfi}{\begin{figure}}
\newcommand{\efi}{\end{figure}}
\newcommand{\ee}{\end{equation}}
\newcommand{\eea}{\end{eqnarray}}
\newcommand{\f}{\frac}
\newcommand{\lbl}{\label}

\newcommand{\bp}{\backprime}
\newcommand{\bpp}{\backprime\backprime}

\newcommand{\vf}{{\bf f}}

\newcommand{\vu}{{\bf u}}

\newcommand{\vrp}{{\bf r}}

\newcommand{\vE}{{\bf E}}

\newcommand{\vF}{{\bf F}}
\newcommand{\vR}{{\bf R}}
\newcommand{\vO}{{\bf O}}

\newcommand{\valpha}{{\bf \mbox{\boldmath {$\alpha$}}}}

\newcommand{\eps}{\epsilon}

\newcommand{\sv}{{\cal V}}

\begin{document}
\setcounter{footnote}{1}

\title{
{\small Absence of a Consistent Classical Equation of Motion for a Mass-Renormalized Point Charge}}
\author{Arthur D. Yaghjian}
%
%
\begin{abstract}
The restrictions of analyticity, relativistic (Born) rigidity, and negligible $O(a)$ terms involved in the evaluation of the self electromagnetic force on an extended charged sphere of radius $a$ are explicitly revealed and taken into account in order to obtain a classical equation of motion of the extended charge that is both causal and conserves momentum-energy.  Because the power-series expansion used in the evaluation of the self-force becomes invalid during transition time intervals immediately following the application and termination of an otherwise analytic externally applied force,  a transition force 
must be included during each of these two transition time intervals to remove the noncausal pre-acceleration and pre-deceleration from the solution to the equation of motion without the transition forces.  Although the exact time dependence of each transition force is not known, the effect of each transition force on the solution to the equation of motion can be determined to within a single unknown constant, the change in velocity of the charge across the transition interval.   For the extended charged sphere, the changes in velocity across the transition intervals can be chosen to maintain conservation of momentum-energy in the causal solutions to the equation of motion within the restrictions of relativistic rigidity and negligible $O(a)$ terms under which the equation of motion is derived. However, regardless of the values chosen for the changes in the velocity across the transition intervals, renormalization of the electrostatic mass to a finite value as the radius of the charge approaches zero introduces a violation of momentum-energy conservation into the causal solutions to the equation of motion of the point charge if the magnitude of the external force becomes too large.  That is, the causal classical equation of motion of a point charge with renormalized mass experiences a high acceleration catastrophe.
\end{abstract}
%
%
\maketitle
\section{\label{sec:Introduction}INTRODUCTION}
The purpose of this paper is twofold: first, to explain why a fully consistent classical equation of motion for a point charge, unlike that of an extended charge, does not exist and, second, to show that this inconsistency is caused by the renormalization of the electrostatic mass of the charge as its radius is allowed to approach zero.  The proof of these results for the mass-renormalized point charge depends critically upon proving the closely related result that the classical equation of motion of the extended charged particle without renormalized mass can be properly modified at its nonanalytic points of time (where the traditional derivation fails) to yield an equation of motion that is causal (free of pre-acceleration and pre-deceleration) and consistent with momentum-energy conservation to within the conditions imposed by the assumptions of relativistic (Born) rigidity and negligible $O(a)$ terms involved in the evaluation of the radiation reaction of the extended charged particle.
%
%
\par
Although the relativistic version of Newton's second law of motion for uncharged particles, and the Maxwell-Lorentz equations for moving charges, are part of the fundamental assumptions of classical physics, it remains uncertain as to how to combine the self electromagnetic force on a moving charge determined from the Maxwell-Lorentz equations with Newton's second law of motion to obtain an equation of motion for a charged particle that obeys both causality and conservation of momentum-energy.  The difficulty lies not only in the impossibility of evaluating the integrals for the self electromagnetic force exactly for all time (because the velocity of the particle is not known \textit{a priori}) but also in not knowing what integral equation, if any, the velocity should satisfy for all time.  \textit{The challenge is much greater than solving a known complicated equation of motion. It is  to examine the self electromagnetic force on a moving charge in hopes of extracting a reasonable classical equation of motion for charged particles that does not violate fundamental principles of classical physics, namely,  causality and conservation of momentum-energy.}
\par
The motivation behind the purely classical approach taken throughout this paper is not to find the equation of motion of an actual fundamental charged particle such as the electron or to find a realistic model for one of these fundamental particles.  The limitations of classical physics imposed by quantum mechanics and quantum electrodynamics are well-known.  However, we can divorce the classical pursuit of a consistent equation of motion from the question of whether or not the idealized model we use for the charge (whose radius may approach zero) approximates an electron (for example) or whether the resultant equation of motion is consistent with quantum physics (even though a robust classical equation of motion may provide insight for its analogue in quantum physics).  We assume that if we could enter an idealized classical laboratory, distribute surface charge uniformly on a perfectly insulating sphere (a continuous medium in which the speed of light is assumed to remain equal to $c$), and apply an external electromagnetic field to the sphere, we would observe a motion that is consistent with causality and momentum-energy conservation and that is predictable by the equations of classical physics.  Unfortunately, if extremely large values of the externally applied force are allowed, no classical equation of motion found to date predicts such a fully consistent motion for a charged particle with renormalized mass as the radius approaches zero --- as will be explained.
\section{\label{ECS}EXTENDED CHARGED SPHERE}
Ultimately, our goal is to obtain a classical equation of motion of a point charge, but since the electrostatic energy of formation and thus the electrostatic mass of a point charge is infinite, it is reasonable to begin with an extended model of a charged particle, namely, an ideal, relativistically rigid, charged insulating sphere (the Lorentz model \cite{Lorentz1892}, \cite{Lorentz1916}). Moreover, the textbook expression for the power radiated by a charge \cite[sec. 14.2]{Jackson} becomes invalid if the velocity or the time derivatives of the velocity of the charge change abruptly with time (such as when the external force is first applied or terminated).    Thus, one begins with an extended charge in hopes of determining an expression for its radiation reaction that remains valid for unrestricted values of the velocity and its time derivatives as the radius of the charge approaches zero.
\par
It is assumed throughout that the sphere is not rotating.  This assumption is justified by the work of Nodvik, who shows that the effect of a finite angular velocity of rotation on the self force and self power of the Lorentz model in arbitrary motion is of $O(a)$ (the order of the terms neglected in the equation of motion of the extended charge),  which approaches zero as the radius of the charge approaches zero \cite[eq. (7.28)]{Nodvik}.  Also, since rotational effects are of $O(a)$ and thus become negligible if $a$ is small enough, the sphere can be assumed to translate without rotation in each instantaneous rest frame.  Thus, the result of von Laue [Phys. Zeit., {\bf 12}, 85 (1911)] that a relativistic body generally has an infinite number of degrees of freedom does not apply to the translating relativistically rigid model \cite{Eriksen-et-al}.  Also, as Pauli [Theory of Relativity, p. 132] explains, although  ``a {\it rigid body} has no place in relativistic mechanics, it is nevertheless useful and natural to introduce the concept of {\it rigid motion} of a body....for which Born's condition is satisfied."
\par
One can immediately postulate an expression for the equation of motion of an extended charge and, in particular, for that of the charged sphere with small radius $a$ and total charge $e$, in the form of the relativistic version of Newton's second law of motion with an added electromagnetic radiation reaction force $\vF_{\rm rad}(t)$ on the charge moving with center velocity $\vu(t)$ (written as just $\vu$)\footnote{Symbols for velocity or its time derivatives written with no expli\-cit functional dependence (for example, $\vu, \dot{\vu}, \ddot{\vu}$) refer to the vel\-ocity or its time derivatives of the center of the spherical particle.} 
\begin{subequations}
\label{1}
\be{1a}
\vF_{\rm ext}(t) + \vF_{\rm rad}(t) = (m_{\rm es} +m_{\rm ins})\frac{d}{dt}(\gamma \vu)\ee 
\be{2a}
\gamma = (1-u^2/c^2)^{-\frac{1}{2}}
\ee
\end{subequations}
where $\vF_{\rm ext}(t)$ is the external force applied to the charged sphere at the time $t$,  $m_{\rm ins}$ is the mass of the uncharged insulator,\footnote{Conceivably, the mass $m_{\rm ins}$ can be negative if it includes the negative formation energies of gravitational or any attractive short-range forces holding the charge together and in place on the insulator \cite{ADM}, \cite{Yaghjian-gravity}.} and $m_{\rm es}$ is the electrostatic mass equal to the  formation energy (needed to bring the charge from infinity to the surface of the sphere) divided by $c^2$
\be{2}
m_{\rm es} = \frac{e^2}{8\pi\eps_0^{} a c^2}
\ee 
with $\eps_0^{}$ the permittivity of free space.
\par
The problem that remains is to determine an expression for the radiation reaction force $\vF_{\rm rad}(t)$ that leads to a consistent equation of motion in (\ref{1a}).  In the following subsection,  this problem is addressed by evaluating the self electromagnetic force $\vF_{\rm em}(t)$ on the moving charged sphere of radius $a$.  Although the self electromagnetic force $\vF_{\rm em}(t)$ includes the force needed to change the electromagnetic momentum of the charged sphere in addition to the radiation reaction force $\vF_{\rm rad}(t)$, this electromagnetic-momentum force can be subtracted at the end of the derivation to yield an expression for $\vF_{\rm rad}(t)$ in terms of the time derivatives of the velocity of the center of the charged sphere.
\subsection{Evaluation of the self electromagnetic force}
In an instantaneous rest frame of the sphere, every point on the sphere moves with the center velocity $\vu(t)=0$ and the self electromagnetic force can be written as a double integral over the charge distribution of the electric-field force between all the differential charge elements on the surface of the sphere \cite[chs. 20--21]{P&P}, \cite[eq. (A.1)]{YB} 
\begin{subequations}
\label{3}
\be{3a}
{\vF}_{\rm em}(t)=\!\!\int\hspace{-3mm}\int\limits_{\mbox{\!\!\!\!\scriptsize{charge}}} \!\!{\vE}[\vrp(t),\vrp'(t'),\vu'(t'),\dot{\vu}'(t')]\, de'de
\ee
with ${\vu}(t)=0$ and ${\vu}'(t)=0$, where $de$ and $de'$ are two differential elements of the surface charge whose positions as a function of time are given by $\vrp(t)$ and $\vrp'(t)$, respectively.   $\vE[\vrp(t),\vrp'(t'),\vu'(t'),\dot{\vu}'(t')]$ is the electric field per unit source charge exerted on $de$ at the present position $\vrp(t)$ due to the  source charge $de'$ at its retarded-time position $\vrp'(t')$ as measured in the reference frame in which the charged sphere is at rest at the time $t$.  It is given  by \cite[chs. 20--21]{P&P}, \cite[eq. (A.2)]{YB}
\bea{3b}
\vE[\vrp,\vrp',\vu',\dot{\vu}']
&=&  \frac{1}{4\pi\epsilon _0\left[1-
\hat{\vR}'\cdot \vu'(t') /c
\right]^3}\left\{\frac{\hat{\vR}'}{R'c^2}\times\right.\nonumber\\&\mbox{}&\hspace{-1.2cm} \left[\left(\hat{\vR}' -\frac{{\vu}'(t')}{c}\right)\times\dot{\vu}'(t')]\right]
\nonumber \\ &\mbox{}& \hspace{-1.2cm} +\frac{1}{R'^2}\left[1-\frac{u'^2(t')}{c^2}\right]\left[\hat{\vR}'-\frac{{\vu}'(t')}{c}\right]\Bigg\}
\eea
where $\vu'(t')= d\vrp'(t')/dt'$ and $\dot{\vu }'(t')=d^2\vrp'(t')/dt'^2$ refer to the
velocity and acceleration of the source charge $de'$ at the retarded time
\be{3c}
t' = t - R'/c
\ee
and the vector $\vR'$ is the difference between the position $\vrp(t)$ of $de$ and the position $\vrp'(t')$ of $de'$ {\em at the retarded time $t'$}
\be{3d}
\vR'=\vrp(t)-\vrp'(t')\,,\;\;\;\;R' =|\vR'|\,.
\ee
\end{subequations}
\par
The integral equation obtained by inserting ${\vF}_{\rm em}(t)$ [minus the electromagnetic  momentum force, $4 m_{\rm es}\dot{\vu}/3$; see (\ref{11a}) and (\ref{14}) below] from (\ref{3a}) into (\ref{1a}) cannot be solved for the center velocity of the sphere because $\vR'$, $\vu'(t')$, and $\dot{\vu}'(t')$ are functions of the position of the charge elements at the retarded time $t'$, for which there is not an explicit expression in terms of the present time $t$.  Moreover, during the time, just after the external force is first applied, that it takes light to traverse the charge distribution, the elecromagnetic momentum force, which must be subtracted from the total self electromagnetic force to get the radiation reaction force, has an unknown value not necessarily equal to $4 m_{\rm es}\dot{\vu}/3$.  Consequently, the usual approach for obtaining a classical equation of motion for a charged particle is to follow the original idea of Lorentz \cite{Lorentz1892}, \cite{Lorentz1916} and derive a power series expansion for the self electromagnetic force with respect to the small radius $a$ of the charged sphere.  For example, in the instantaneous rest frame of time $t$, the derivatives of velocity expand in a Taylor series as
\begin{subequations}
\label{4}
\be{4a}
{\vu}'(t')=
-\dot{{\vu}}'(t)\frac{R'(t')}{c}+\ddot{{\vu}}'(t)
\frac{R'^2(t')}{2c^2}+ \cdots  
\ee
\be{4b}
\dot{\vu}'(t')=
\dot{{\vu}}'(t)-\ddot{{\vu}}'(t)\frac{R'(t')}{c}
+ \cdots  
\ee
where the distance $R^\prime (t^\prime )$ has the Taylor series expansion
\be{4c}
R'(t')=R(t)-\frac{R(t){\vR}\cdot\dot{{\vu}}'(t)}{2c^2}+ \cdots\;.
\ee
\end{subequations}
\subsubsection{Validity of the Taylor series expansions}
These Taylor series expansions in (\ref{4}) are valid  {\em provided the velocity
function $\vu'(t')$ is an analytic function of complex time $t'$ for}
\be{5}
|t' -t|\le [R'(t')/c]_{\mbox{\scriptsize max}}\,.
\ee
(Analyticity of $\vu'(t')$ implies the analyticity of $R'(t')$ and $\dot{\vu}'(t')$ through integration and differentiation, respectively.)
For the self-force calculation in the rest frame,
$R^\prime (t^\prime )$ does not exceed a value of about $2a$
(assuming the velocity does not change rapidly between
$t^\prime$ and $t$; in other words, assuming the velocity change is
a small fraction of the speed of light during the time it takes light
to traverse the charge distribution), and thus (\ref{5}) can be rewritten as
\begin{subequations}
\label{6}
\be{6a}
|t' -t|\le \Delta t_a
\ee
where $\Delta t_a \approx {2a/c}$.  Even if the magnitude of the velocity change ($\Delta u$) during the time $2a/c$ is a significant fraction of the speed of light, we have
\be{6b}
\Delta t_a \approx \f{2a}{c} \f{1}{1-|\Delta u| /c} =O(a/c)\,.
\ee
\end{subequations}

\par
Suppose, for example, that an externally applied force that is zero for $t<t_1$ turns on at $t=t_1$ and is an analytic function of time in a complex neighborhood of the real $t$ axis until it turns off at $t = t_2$, after which time it remains zero.  Then the Taylor series in (\ref{4}) hold for all  $-\infty < t < +\infty$ except in the intervals 
\be{7}
 t \in [t_1, t_1 +\Delta t_{a1}]\;\;\mbox{  and  }\;\; t \in [t_2, t_2 +\Delta t_{a2}]
\ee
where $\Delta t_{a1}$ and $\Delta t_{a2}$ are transition time intervals of $O(a/c)$. Physically, $\Delta t_{a1}$ and $\Delta t_{a2}$ are the times  it takes immediately after the external force is first applied and immediately after the external force is removed, respectively,  for the abrupt change in radiation from each element of charge to communicate itself to all the other elements of charge on the sphere.  (The $\Delta t_{a}$'s given in (\ref{6b}) and (\ref{7}) are for the instantaneous rest frames just before the changes in external forces occur.  However, since they remain of $O(a/c)$ in every other inertial frame including the laboratory frame, throughout the paper we shall denote these transition time intervals by the same symbol, $\Delta t_{a}$, for all inertial reference frames.)
\par
Abraham realized that the traditional series representation of
the self electromagnetic force became invalid for ``discontinuous
movements" of the charge.  In \cite[sec. 23]{r:2} he states, ``These two
forces [electromagnetic momentum term plus radiation reaction] are
basically nothing other than the first two terms of a progression
which increases in accordance with increasing powers of the
electron's radius $a$. $\ldots $  Because the internal force is
determined by the velocity and acceleration existing in a finite
interval preceding the affected point in time, such a progression is
always possible when the movement is continuous and its velocity is
less than the speed of light. $\ldots $  The series will converge
more poorly the closer the movement approaches a discontinuous
movement and the velocity approaches the speed of light . $\ldots $
It fails completely for discontinuous movements. $\ldots $ Here,
other methods must be employed when computing the internal force."
Abraham goes on to derive the radiated energy and momentum of a
charged sphere with discontinuous velocity \cite[sec. 25]{r:2}, \cite{r:39}.  He also derives
Sommerfeld's general integral formulas for the internal
electromagnetic force \cite{r:40}.  Neither he nor Sommerfeld, however,
evaluates or interprets these general integrals except to show they
yield a null result for a charged sphere moving with constant
velocity.
\par
Schott \cite{Schott-1908}, \cite[p. 283]{r:13} also concludes that ``the approximation [used to obtain the Lorentz-Abraham equation of motion] fails during an interval of time, which is comparable with the time required by an electromagnetic wave to pass across the electron and includes the instant at which the discontinuity occurs."
\par
More recently, Valentini \cite{Valentini} observes that ``the usual derivations of the Lorentz-[Abraham-]Dirac equation are only valid at times such that [the position of and force applied to the particle] are analytic functions [of time]," and that nonanalyticity of these functions is responsible for the noncausal pre-acceleration in the solution to the Lorentz-Abraham-Dirac equation of motion.  However, the modified solution proposed by Valentini did not take into account changes in velocity across the transition intervals (see Section III.A below) and thus violated conservation of energy \cite{Goebel}.
\subsubsection{Surface-charge accelerations in terms of center acceleration of sphere (requirements of relativistic rigidity)}
The acceleration $\dot{\vu}'(t)$ and its derivatives in (\ref{4}) at the present time $t$ are those of the charge elements $de'$ on the surface of the sphere in the instantaneous inertial rest frame of the charged sphere at time $t$.  Because of the Lorentz contraction of the sphere,  the values of these accelerations and their time derivatives on the surface of the sphere are different from the corresponding values for the center of the sphere.  Therefore, before substituting from (\ref{4}) into (\ref{3b}), we need to determine the values of  $\dot{{\vu}}'(t)$ and $\ddot{{\vu}}'(t)$ in terms of the center values denoted by $\dot{{\vu}}(t)=\dot{{\vu}}$ and $\ddot{{\vu}}(t)=\ddot{{\vu}}$.
\par
The charged sphere is assumed to be relativistically rigid in the sense that the relative position of each material point of the nonrotating sphere remains the same in every instantaneous inertial rest frame.  Thus, the problem of finding the velocity, acceleration, and higher derivatives of velocity of each point of the surface of the sphere in terms of the corresponding values of the center of the sphere is identical to the problem of relativistic rigidity first proposed and studied by Born \cite{Born}. Specifically, as the radius $a$ of the sphere gets small, it can be shown (see \cite[eq. (A.3)]{Newman&Janis})\footnote{There is a factor of $\ddot{f}(t)$ (equal to our $\dot{u}(t)$) missing in the third term on the right-hand side of (A.3) in \cite{Newman&Janis}.} that in the instantaneous rest frame ($\vu (t) = 0$) \cite[eqs. (A.8) and (A.9)]{YB}
\begin{subequations}
\label{8}
\be{8a}
\dot{{\vu}}'(t)=\dot{{\vu}}(t)-\frac{{\vrp}'(t)\cdot\dot{{\vu}}(t)}
{c^2}\dot{{\vu}}(t)+\vO\left(a^2\right)
\ee
\be{8b}
\ddot{\vu}'(t) = \ddot{\vu}(t) + \vO(a)
\ee
\end{subequations}
where $\vrp'(t)$ is the position (measured from the center of the sphere) of $de'$ on the surface of the sphere in the rest frame at time $t$.  The symbol $\vO(a^m)$ means $\sum_{n=m}^{\infty} \valpha_n (\vu) a^n$ where the $\valpha_n (\vu)$ are finite functions of velocity (and its time derivatives) but not functions of $a$.  By applying the results derived in \cite{Eriksen-et-al}, it is found in Appendix A that the relations in (\ref{8}) hold under the restriction that
\be{9}
|\dot{\vu}| \ll \f{c^2}{a}\,.
\ee  
Moreover, it is shown in Appendix A that if  the center acceleration is as large as $|\dot{\vu}| = {c^2}/{a}$, then the acceleration of the end of the sphere in the direction opposite the acceleration becomes infinite while the acceleration of the other end of the sphere has an acceleration equal to one half the  center acceleration.  The velocity across the sphere also varies greatly and nonlinearly from its center value.   {\em Thus, unless the inequality in (\ref{9}) is satisfied in the instantaneous rest frame, it becomes impossible to describe the motion of all the charge on the sphere by the motion of the center of the sphere (or by the motion of any other single point of the sphere).}
\par
Substituting (\ref{8}) into (\ref{4}), the resulting equations into (\ref{3b}), then $\vE[\vrp,\vrp',\vu',\dot{\vu}']$ into (\ref{3a}) yields \cite[eq. (A.10)]{YB}
\bea{10}
&&{\vF}_{\rm em}(t)=\f{1}{4\pi\eps_0^{}}\int\hspace{-3mm}\int\limits_{\mbox{\!\!\!\!\scriptsize{charge}}}\!\!\left\{ \frac{\hat{{\vR}}}{R^2}+\frac{1}{2c^2 R}\left[ \frac{{\vrp}'\cdot
\dot{{\vu}}}{c^2}-1\right]\hspace{10mm}\right.\nonumber\\
&&\left.\hspace{10mm}\cdot\left[(\hat{{\vR}}\cdot\dot{{\vu}})\hat{{\vR}}+
\dot{{\vu}}\right]+\frac{3}{8}\frac{\hat{{\vR}}}{c^4}
\left[(\hat{{\vR}}\cdot\dot{{\vu}})^2 -|\dot{{\vu}}|^2\right]\right.\nonumber\\       &&\left.\hspace{18mm}+\frac{3(\hat{{\vR}}\cdot\dot{{\vu}})\dot{{\vu}}}{4c^4} +\frac{2\ddot{{\vu}}}{3c^3}+\vO(a)\right\} de'de 
\eea
with $\vR  = \vrp(t)  - \vrp '(t)$, and $\vu(t) = 0.$   All the terms with
an odd number of products of $\hat{\vR }$ or $\vrp '$ integrate to
zero and the remaining even product terms integrate to give the
well-known expression for the self electromagnetic force in the instantaneous rest
frame
\be{11a}
{\vF}_\mathrm{em }(t)=-\frac{e^2}{6\pi\epsilon _0 ac^2}
\dot{{\vu}}+\frac{e^2}{6\pi\epsilon _0 c^3}
\ddot{{\vu}}+\vO(a),\;\;\; u=0\;.
\ee 
By carefully going through the derivation, however, we have revealed two important restrictions on the validity of (\ref{11a}).  First, as explained in Section II.A.1, the derivation of (\ref{11a}) requires local analy\-ticity with time of the velocity and thus of the externally applied force.   For an external force that is an analytic function of time in a neighborhood of the real time axis, except for when it turns on at $t = t_1$ and when it turns off at $t=t_2$, (\ref{11a}) holds for all time $t$ except during the $O(a/c)$ transition intervals that occur immediately after $t_1$ and $t_2$ and that are given in (\ref{7}).  Second, as explained above in this section, the requirement of relativistic rigidity invalidates the derivation of (\ref{11a}) if the limitation on the magnitude of the acceleration in (\ref{9}) is not satisfied.  Consequently, (\ref{11a}) must be qualified by the rest-frame conditions
\be{11b}
t \notin [t_1, t_1+\Delta t_{a1}]\,,\;\;\;\;\;t \notin [t_2, t_2+\Delta t_{a1}]
\ee
and
\begin{subequations}
\label{11rigid}
\be{11c}
|\dot{\vu}| \ll \f{c^2}{a}\,,\;\;\;\;  -\infty < t < +\infty.
\ee
Integrating the inequality in (\ref{11c}) over a transition interval $\Delta t_a$ given in (\ref{6b}) for a rest frame at the beginning of the transition interval shows that $|\Delta u|(1 -|\Delta u|/c) \ll 2c$ or
\be{11d}
\f{|\Delta u|}{c} \ll 1
\ee
\end{subequations}
where $\Delta u$ is the velocity change across the rest-frame transition interval.
\par
It can also be shown that for the $\vO(a)$ terms in (\ref{11a}) to be negligible, the following conditions must be satisfied in the rest frame for $t$ outside the transition intervals given in (\ref{11b}) \cite[p. 74]{YB}
\begin{subequations}
\label{12}
\be{12a}
|\dot{\vu}| \ll \f{c^2}{a}
\ee
\be{12b}
\frac{c}{a}\left|\sum_{n=2}^{\infty}\left(\f{-2a}{c}\right)^n\f{1}{n!}\frac{d ^{n}\vu}{d t^{n}}\right| \ll \left( \left|\frac{d \vu}{d t}\right|, \f{c^2}{a}\right)\, .
\ee
\end{subequations}
\par
The electromagnetic momentum of a charged sphere of small radius $a$ moving with speed $u\ll c$ is given by $m_{\rm em}\vu$, where the ``electromagnetic mass"  is given by  \cite[secs. 21-4 and 21-5]{P&P}\footnote{The 4/3 factor in the electromagnetic mass has been the subject of discussion in many publications since the time of Lorentz and Abraham, although Lorentz and Abraham were unconcerned with this factor in their original work because it was done before Einstein's 1905 papers on relativistic electrodynamics and the mass-energy relation.  (They were concerned, however, with the  self electromagnetic power expression corresponding to (\ref{11a})  not agreeing in the first term with the result of taking the dot product of $\vu$ with (\ref{11a}) --- a discrepancy that was removed by Poincare's determination of the power contributed by the forces that bond the charge to the sphere \cite{YB}.)  From a fundamental perspective, the electromagnetic mass need not equal the electrostatic mass because the electromagnetic stress-momentum-energy tensor is not divergenceless in the presence of charge-current and thus the associated electromagnetic momentum-energy will not generally transform as a relativistic four-vector \cite{Schwinger}.} 
\be{13}
m_{\rm em} =\f{4}{3} m_{\rm es} = \frac{e^2}{6\pi\eps_0^{} a c^2}\,.
\ee
Thus, the $\dot{\vu}$ term in (\ref{11a}) is the force in the rest frame required to change the electromagnetic momentum of the charged sphere.  It must be removed from the self electromagnetic force to obtain the radiation reaction force; that is 
\be{14}
{\vF}_\mathrm{rad}(t)=
\frac{e^2}{6\pi\epsilon _0 c^3}
\ddot{{\vu}}+\vO(a),\;\;\; u=0\,.
\ee 
Insertion of the radiation reaction force from (\ref{14}) into (\ref{1a}) produces the following equation of motion in the instantaneous rest frame
\be{15}
\vF_{\rm ext}(t)  = (m_{\rm es} +m_{\rm ins})\dot{\vu} - \frac{e^2}{6\pi\epsilon _0 c^3}
\ddot{{\vu}}+\vO(a),\;\;\; u=0
\ee 
{\em under the restrictions given in (\ref{11b}) and (\ref{11rigid})}.
\section{CAUSAL EQUATION OF MOTION HOLDING FOR ALL TIME}
For an external force that is an analytic function of time between the time it turns on at $t=t_1$ and turns off at $t=t_2$, the rest-frame equation of motion in (\ref{15}) holds for all $-\infty < t < +\infty$ except in the time intervals given in (\ref{11b}) just after the external force turns on and just after it turns off.  During these transition time intervals, we cannot evaluate the self electromagnetic force in (\ref{3a}) because the velocity of the charged sphere is an unknown, possibly rapidly varying function of time during these short time intervals.  Also, the electromagnetic self force may contain delta-like functions and their derivatives in these short time intervals because the highly singular $1/R'^2$ fields in (\ref{3b}) contribute to the integral in (\ref{3a}) during these intervals.
\par
Although we do not know the form of the equation of motion during these transition intervals, if an equation of motion exists for all time, it must equal some function during these time intervals.  That is, assuming a consistent classical equation of motion exists for the center velocity of the charged sphere at all times, we can express it as \cite[sec. 8.2.2]{YB}
\be{16}
\vF_{\rm ext}(t)+ \vf_{a1}(t) +\vf_{a2}(t) = (m_{\rm es} +m_{\rm ins})\dot{\vu} - \frac{e^2}{6\pi\epsilon _0 c^3}\ddot{{\vu}}+\vO(a) 
\ee 
{\em where $\vf_{a1}(t)$ and $\vf_{a2}(t)$ are unknown transition self forces that are zero outside their respective intervals given in (\ref{11b}) and that may contain delta functions and their derivatives  as $a \to 0$.  The equation of motion in (\ref{16}) holds for all $-\infty < t< +\infty$ under the restrictions in (\ref{11rigid}) imposed by relativistic rigidity on the magnitude of the acceleration and on the velocity changes across the transition intervals.}
\par
If the conditions in (\ref{12}) for neglecting the $\vO(a)$ terms are satisfied, the rest-frame equation of motion in (\ref{16}) becomes 
\be{17}
[{\vF_{\rm ext}(t)+ \vf_{a1}(t) +\vf_{a2}(t)}]/{m} = \dot{\vu} - \tau_e\ddot{{\vu}}
\ee
with 
\be{18}
m = m_{\rm es} + m_{\rm ins}
\ee
and
\be{19}
\tau_e = \f{e^2}{6\pi\eps_0^{}mc^3}\,.
\ee
The rest-frame equation of motion in (\ref{17}) transforms to an arbitrary inertial frame of reference as \cite[eq. (8.45a)]{YB}
\begin{subequations}
\label{20}
\bea{20a}
&&\hspace{-5mm}\f{\vF_{\rm ext}(t)+ \vf_{a1}(t) +\vf_{a2}(t)}{m} = \f{d(\gamma \vu)}{dt} -  \tau_e\left\{ \f{d }{d t} \left[\gamma \frac{d }{d t}(\gamma\vu) 
\right]\right. \nonumber\\  
&&\left.\hspace{26mm} - \f{\gamma^4}{c^2} 
\left[|\dot{\vu}|^2 +\f{\gamma^2}{c^2}(\vu\cdot\dot{\vu})^2 \right] 
\vu \right\}
\eea
or in four-vector notation \cite[eq. (8.168)]{YB}
\be{20b}
\f{F_\mathrm{ext}^i\! +\! f_{a1}^i\! + \!f_{a2}^i}{mc^2} = \frac{d u^i}{d s}
-\tau_e\left(\frac{d ^2 u^i}{d s^2}+ u^i\frac{d u_j}{d s}\frac{d u^j}{d s} 
\right)
\ee
\end{subequations}
provided the conditions in (\ref{12}) for neglecting the $O(a)$ terms are satisfied outside the transition intervals in (\ref{11b}) and the relativistic rigidity condition in (\ref{11c}) is satisfied for all $t$, including times within the transition intervals, so that (\ref{11d}) is satisfied across each transition interval.  Herein, the dimensionless four-vector notation of Panofsky and Phillips \cite{P&P} is used, where $u^i = \gamma(\vu/c,1)$, $u_i =\gamma(-\vu/c,1)$, and $ds = c\, dt/\gamma$.
\subsection{Causal solutions to the equation of motion: elimination of the pre-acceleration and pre-deceleration}
Although the exact values of the transitional self forces $\vf_{a1}(t)$ and $\vf_{a2}(t)$ are unknown (because of the unknown time dependence of the velocity across the transition intervals), remarkably, they can be chosen to completely eliminate the pre-acceleration and pre-deceleration from the solutions to the original equation of motion without these transitional self forces.  This result is proven in \cite[sec. 8.2.2]{YB} for the general equation of motion in (\ref{20}), but here we shall concentrate on rectilinear motion for which (\ref{20a}) simplifies to
\bea{21}
\frac{F_\mathrm{ext}(t)+f_{a1}(t) +f_{a2}(t)}{m} &=&  \f{d (\gamma u)}{d t} -
\tau_e\left\{ \f{d }{d t} \left[\gamma \frac{d }{d t}(\gamma u) 
\right]\right.\nonumber\\
&&\hspace{10mm}-\left.  \f{\gamma^6}{c^2} \dot{u}^2 u \right\}.
\eea
The substitutions $dt = \gamma d\tau$ and $\gamma u/c = \sinh(\sv/c)$ convert this nonlinear equation of rectilinear motion to the linear equation of motion
\be{22}
\f{F_\mathrm{ext}(\tau)+f_{a1}(\tau)+f_{a2}(\tau)}{m}=\sv^\bp(\tau)-\tau_e\sv^{\bpp}(\tau)
\ee
where the backprimes indicate differentiation with respect to the proper time $\tau$ and it is assumed that the conditions in (\ref{12}) are satisfied so that the $O(a)$ terms are negligible.
\par
It is shown in \cite[sec. 8.2.3]{YB} that the transition self forces for the rectilinear equation of motion in (\ref{22}) can be expressed as
\bea{23}
&&\hspace{-4mm}\frac{f_{an}(\tau)}{m} = [\Delta \sv_n -\tau_e \Delta\sv^\bp_n]\,
\delta(\tau-\tau_n^+) -\tau_e \Delta\sv_n \,\delta^\bp(\tau -\tau_n^+)\nonumber\\
&&\hspace{-3mm}n=1,2
\eea 
where $\Delta \sv_n$ and $\Delta\sv^\bp_n$ are the jumps in $\sv$ and $\sv^\bp$ across the two short transition intervals of duration $\Delta t_{an}$ for small $a$ and $\tau_n,\; n=1,2$, are the proper times at which the external force turns on and off, respectively.  The superscript $^+$ on $\tau_n^+$ indicates a time between $\tau_n$ and $\tau_n +\Delta t_{an}$, and for a finite (nonzero) value of $a$, the delta functions can be considered to be finite in height and spread out across the transition intervals.
 $\Delta \sv^\bp_n$ is determined solely by the externally applied force and is independent of $f_{an}(\tau)$.  Thus, it is a parameter whose value cannot be changed in (\ref{23}).  However, $f_{an}(\tau)$ alone determines $\Delta \sv_n$ and thus we are free to decide the value of $\Delta \sv_n$ in (\ref{23}).  Choosing $\Delta \sv_n =0$ leaves only the delta function in (\ref{23}) and makes the velocity function continuous.  Choosing $\Delta \sv_n =\tau_e \Delta\sv^\bp_n$ leaves only the doublet function in (\ref{23}) and produces a jump in velocity approximately equal to the change in velocity produced by the pre-acceleration or pre-deceleration in the equation of motion without the transition self forces. (In Section III.B below, it is shown that the jumps in velocity across the transition intervals cannot be chosen arbitrarily if energy-momentum is to be conserved and, moreover, that these jumps in velocity are determined simply in terms of the jumps in acceleration across the transition intervals if the charged sphere moves to minimize the energy radiated during the transition intervals.)
\par
The delta-like and doublet-like functions in (\ref{23}) allow the magnitude and direction of a transition force to change dramatically over its transition interval. Such dramatic changes are compatible with contributions from the self-force integral in (\ref{10}) when the velocity and its time derivatives are changing rapidly during the transition intervals following nonanalytic points in time of the externally applied force.
\par
The solution to (\ref{22}) with the transition forces in (\ref{23}) and with the velocity of the sphere zero before $\tau_1 =0$ is given outside the transition intervals by \cite[eqs. (8.56) and (8.72b)]{YB}
\begin{subequations}
\label{24}
\bea{24a}
&&\hspace{-9mm}\sv^\bp(\tau)=\frac{1}{m\tau_e}\nonumber\\ &&\hspace{-10mm}\cdot\left\{\begin{array}{lll} 0\!&,&\!\tau < 0\\
\!\int\limits_\tau^\infty F_1(\tau_0)\exp[-(\tau_0-\tau)/\tau_e] d \tau_0\!&,&\!\Delta t_{a1} < \tau < \tau_2\\[3.mm]
0\!&,&\!\tau_2+\Delta t_{a2} < \tau
\end{array}\right.
\eea 
\bea{24b}
&&\sv(\tau) =\tau_e \sv^\bp(\tau)+\sum_{n=1}^2 h(\tau -\tau_n) 
(\Delta \sv_n -\tau_e \Delta \sv^\bp_n)\nonumber\\&&\hspace{-9mm}+\frac{1}{m}\int\limits_{0}^\tau F_\mathrm{ext}(\tau_0) d \tau_0\,,\;\tau \notin \{[0,\Delta t_{a1}], [\tau_2, \tau_2 +\Delta t_{a2}]\}  
\eea
\end{subequations}
where $h(\tau)$ is the unit step function and $F_1(\tau)$ in (\ref{24a}) is the analytic continuation of the external force $F_{\rm ext}(\tau)$ from $\tau < \tau_2$ to  $\tau \ge \tau_2$.  The jumps $\Delta \sv^\bp_n$ in (\ref{24b}) across the two transition intervals can be found in terms of $F_1(\tau)$ from (\ref{24a}).
\par
Although the solution in (\ref{24a}) is
free of pre-acceleration and pre-deceleration, it may be bothersome that for $\Delta t_{a1}< \tau < \tau_2$ the solution in (\ref{24a}) to the equation of motion depends on the values of the analytically continued external force at all future times.  This result becomes understandable if it is remembered that (\ref{24a}) is the solution to an equation of
motion obtained under the restriction that the analytically continued externally applied
force function $F_1(\tau)$ be an analytic function of time about the real $\tau$ axis for all $\tau > 0$, because the values of an analytic function on an interval of a singly connected domain of analyticity determine uniquely the function over the rest of the domain.  For example, assume that for $\tau > 0$ the external force 
$F_1(\tau_0)$ in (\ref{24a}) can be expanded in a power
series about $\tau$ to recast (\ref{24a}) in the form
\be{24'}
\sv^\bp(\tau )=\frac{1}{m}
\sum_{k=0}^{\infty}(\tau_e)^k\frac{
d ^k F_n(\tau )}{d \tau^k},
\;\;\;\;  \Delta t_{a1} < \tau <\tau_2
\ee   
which simply states that the acceleration at any one time $(\Delta t_a < \tau <\tau_2)$ depends on the time derivatives of
the applied force as well as the applied force itself at that time.
(Note that (\ref{24'}) is not a valid representation for $\sv^\bp(\tau)$ in the transition interval $0 < \tau < \Delta t_{a1}$ containing the transition force in addition to the externally applied force.)
\par
If the restriction that the analytically continued external force $F_1(\tau)$ be an analytic function of $\tau$ for all $\tau >0$ is ignored, and $F_1(\tau)$ is allowed to attain a strong enough infinite singularity at some future point in time, as in the case of the charged sphere being attracted to the center of a Coulomb field ($1/r^2$ singularity), the integration in (\ref{24a}) may not converge for all values of $\tau$ before the sphere reaches the singularity \cite{Baylis-singularity}.
\subsubsection{Charge in a uniform electric field for finite time}
The rectilinear solution in (\ref{24}) takes an especially simple form if the charged sphere is accelerated by a uniform electrostatic field $E_0$ for a finite time from $t_1 = \tau_1 = 0$ to $t = t_2$ ($\tau = \tau_2$).  For example, the charge could be accelerated between two infinitesimally thin plates of a parallel-plate capacitor charged to produce the electric field $E_0$.  It could be released at time $t=0$ from one plate of the capacitor and leave through a small hole in the second plate at time $t=t_2$. Then (\ref{24}) become
\begin{subequations}
\lbl{25}
\be{25a}
\sv^\backprime(\tau)=\frac{eE_0}{m}\left\{\begin{array}{lll}0 & ,\;\; & \tau < 0\\[2mm]
1 & ,\;\; & 
\Delta t_{a1} <\tau < \tau_2\\[2mm]
0 & , \;\;&\tau_2 +\Delta t_{a2} < \tau
\end{array}\right.
\ee
\be{25b}
\sv(\tau)=\left\{\begin{array}{lll} 0 & ,\;\; & \tau < 0\\[2mm]
\Delta \sv_1 + eE_0\tau /m & ,\;\; & 
\Delta t_{a1} < \tau < \tau_2\\[2mm]
\Delta \sv_{21} +eE_0\tau_2 /m & , \;\;&\tau_2 +\Delta t_{a2} < \tau
\end{array}\right.
\ee
\end{subequations}
with $\Delta \sv_{21}=\Delta \sv_{2}+\Delta \sv_{1}$.  These equations recast in terms of $u(t)$ as
\begin{subequations}
\lbl{26}
\be{26a}
\frac{d (\gamma u)}{d t} = \gamma^3 \dot{u}=\frac{eE_0}{m}\left\{\begin{array}{lll}0 & ,\;\; & t < 0\\[2mm]
1 & ,\;\; & 
\Delta t_{a1} <t < t_2\\[2mm]
0 & , \;\;&t_2+ \Delta t_{a2}< t
\end{array}\right.
\ee
\be{26b}
\gamma u=\left\{\begin{array}{lll} 0 & , \;\;& t < 0\\[2mm]
\Delta(\gamma u)_{_{\mbox{\scriptsize 1}}} + eE_0 t/m & , \;\;& 
\Delta t_{a1} < t < t_2\\[2mm]
\Delta(\gamma u)_{_{\mbox{\scriptsize{21}}}} +eE_0 t_2 /m & ,\;\; &t_2+ \Delta t_{a2}< t
\end{array}\right.
\ee
with $\gamma$ found from $\gamma u$ by the relation
\be{26c}
\gamma(t) = \left\{1 + [\gamma(t)u(t)/c]^2 \right\}^{1/2}
\ee
\end{subequations}
and $\Delta(\gamma u)_{_{\mbox{\scriptsize{21}}}} =\Delta(\gamma u)_{_{\mbox{\scriptsize 2}}}+ \Delta(\gamma u)_{_{\mbox{\scriptsize 1}}}$, where $\Delta(\gamma u)_{_{\mbox{\scriptsize 1}}}$ and $\Delta(\gamma u)_{_{\mbox{\scriptsize 2}}}$ are the jumps in $\gamma(t) u(t)$ across the transition intervals at $t=t_1=0$ and $t=t_2$.
\par
In contrast to these causal solutions to the equation of motion in (\ref{21})--(\ref{22}) with the transition self forces, the solution to the equation of motion without these transition forces exhibit pre-acceleration and pre-deceleration.  For example, the solution to (\ref{22}) without the transition self forces for the charge moving through the uniform electric field of a parallel-plate capacitor is given by
\begin{subequations}
\lbl{27}
\be{27a}
\sv^\backprime_{\mbox{\scriptsize pre}}(\tau) =\frac{eE_0}{m}\left\{\begin{array}{lll}\left(1-e^{-\tau_2/\tau_e}\right)e^{\tau/\tau_e} & ,\;\; & \tau \le 0\\[2mm]
\left(1-e^{(\tau -\tau_2)/\tau_e}\right) & ,\;\; & 
0 \le \tau \le \tau_2\\[2mm]
0 & ,\;\; &\tau_2 \le \tau
\end{array}\right.
\ee
\be{27b}
\sv_{\mbox{\scriptsize pre}}(\tau) =\frac{eE_0}{m}\left\{\begin{array}{lll}\tau_e\left(1-e^{-\tau_2/\tau_e}\right)e^{\tau/\tau_e} & ,\;\; & \tau \le 0\\[2mm]
\tau_e\left(1-e^{(\tau -\tau_2)/\tau_e}\right) + \tau & , \;\;& 
0 \le \tau \le \tau_2\\[2mm]
\tau_2 & ,\;\; &\tau_2 \le \tau\,.
\end{array}\right.
\ee
\end{subequations}
\par
One sees from this example of the motion of a charge through a parallel-plate capacitor that the transition forces $f_{an}(t)$, which are nonzero only during the short time intervals following the points in time where the externally applied force is discontinuous, remove both the noncausal pre-\-ac\-cel\-er\-a\-tion and pre-\-de\-cel\-er\-a\-tion from the solution to the equation of motion.  However,  the transition forces $f_{an}(t)$ in the equation of motion change, in general, the momentum and energy of the charged sphere \cite{Baylis}. The next section determines conditions under which this change in momentum-energy is consistent with the conservation of momentum-energy and a non-negative radiated energy during the transition intervals.
\subsection{Conservation of momentum-energy in the causal equation of motion}
The transition forces ensure that the solutions to the equation of motion in (\ref{20}) or (\ref{21})--(\ref{22}) obey causality while remaining free of runaway motion.  However, these transition forces, in general,  change the momentum and energy of the charged particle. 
Consider, for example, the power equation of rectilinear motion obtained by multiplying (\ref{21}) by $u$
\bea{28}
\frac{[F_\mathrm{ext}(t)+f_{a1}(t)+f_{a2}(t)]u}{mc^2}  &=& \f{d \gamma}{d t} -\tau_e
 \left[ \f{d }{d t}\left(\gamma\frac{d \gamma}{d t}\right)\right.\nonumber\\
&&\hspace{6mm}-\left. \frac{\gamma^6}{c^2} \dot{u}^2 \right]. 
\eea
Integrating this power equation of motion from the time $t=t_1=0$ before the external force is first applied and the velocity of the charge is zero to a time $t>t_1$ gives
\bea{29}
&&\hspace{-9mm}\frac{1}{mc^2} \int\limits_{0}^{t} F_\mathrm{ext}u\, dt =\left[\gamma(t)-1\right] -\tau_e\gamma(t)\dot{\gamma}(t)\nonumber\\
&&\hspace{-8mm}+ \frac{1}{mc^2} \int\limits_{0}^{t} m\tau_e \gamma^6 \dot{u}^2dt-\frac{1}{mc^2} \int\limits_{0}^{t}\left[f_{a1}+f_{a2}\right]u\, dt.
\eea
Between the time $t=0$ and the time $t= t_2^+=t_2 + \Delta t_{a2}$, a time  $\Delta t_{a2}$ after the external force has turned off, there appears to be no reason why the energy from the transition forces cannot contain both reversible and irreversible (radiated energy) contributions.  After the time $t=t_2^+$ that the external force is turned off, (\ref{29}) becomes
\bea{30}
&&\hspace{-9mm}\frac{1}{mc^2} \int\limits_{0}^{t_2^+} F_\mathrm{ext}u\, dt =\left[\gamma(t_2^+)-1\right] \nonumber\\
&&\hspace{-8mm}+ \frac{1}{mc^2} \int\limits_{0}^{t_2^+} \left[m\tau_e \gamma^6 \dot{u}^2-\left(f_{a1}+f_{a2}\right)u\right] dt.
\eea
The integral on the left-hand side of (\ref{30}) is the total work done by the external force and the first term (in square brackets) on the right-hand side of (\ref{30}) is the kinetic energy (divided by $mc^2$) of the nonradiating charged sphere moving with constant velocity after the external force has been turned off.  By the Einstein mass-energy relationship, this kinetic energy of the nonradiating charged sphere moving with constant velocity is its total change in energy from its original rest energy.  Thus, the integral on the right-hand side of (\ref{30}) is the total energy radiated by the charged sphere. (Recall that the energy radiated by an extended charge whose velocity changes abruptly during the time it takes light to traverse the charge, that is, during the transition time intervals, is not given by just the integral of $m\tau_e \gamma^6 \dot{u}^2$.)
\par
If $|F_{\rm ext}(t)| \Delta t_{an}/(mc)\ll 1$ for $t\in [t_{an}, t_{an}+\Delta t_{an}], n=1,2$  (conditions that are always satisfied by finite external forces as $a \to 0$),  the integral of the external force over the transition intervals is negligible and (\ref{30}) can be rewritten as
\bea{31}
&&\hspace{-9mm}\frac{1}{mc^2} \int\limits_{t_1^+}^{t_2} F_\mathrm{ext}u\, dt =\left[\gamma(t_2^+)-1\right] \nonumber\\
&&\hspace{-8mm}+ \frac{1}{mc^2} \int\limits_{0}^{t_2^+} \left[m\tau_e \gamma^6 \dot{u}^2-\left(f_{a1}+f_{a2}\right)u\right] dt
\eea
where $t_1^+ = \Delta t_{a1}$.
With the integral of $F_{\rm ext}u$ in (\ref{31}) over the time between the two transition intervals given from (\ref{28}) as
\bea{32}
&&\hspace{-9mm}\frac{1}{mc^2} \int\limits_{t_1^+}^{t_2} F_\mathrm{ext}u\, dt = -\tau_e\left[\gamma(t_2)\dot{\gamma}(t_2)- \gamma(t_1^+)\dot{\gamma}(t_1^+)\right]\nonumber\\
&&+\left[\gamma(t_2)-\gamma(t_1^+)\right] +\frac{1}{mc^2} \int\limits_{t_1^+}^{t_2} m\tau_e \gamma^6 \dot{u}^2 dt
\eea
(\ref{31}) yields
\bea{33}
\f{W_{TI}}{mc^2} &=&\frac{1}{mc^2}\int\limits_{\mbox{\scriptsize TIs}} \left[m\tau_e \gamma^6 \dot{u}^2-\left(f_{a1}+f_{a2}\right)u\right] dt \nonumber\\ 
&=&-\tau_e\big[\gamma(t_2)\dot{\gamma}(t_2) - \gamma(t_1^+)\dot{\gamma}(t_1^+)\big]\nonumber\\
&&-\left[\gamma(t_2^+)-\gamma(t_2) + \gamma(t_1^+)-1\right] 
\eea
where the abbreviation ``TIs" on the integral sign stands for ``transition intervals."  Since this integral on the left-hand side of (\ref{33}) is the energy radiated ($W_{TI}$) by the charged sphere during the two transition intervals, it must be equal to or greater than zero.  Thus, the rectilinear equation of motion in (\ref{21}) is consistent with energy conservation only if the jumps in velocity across the transition intervals can be chosen to make the right-hand side of (\ref{33}) equal to or greater than zero.
\par
The jumps in velocity across the transition intervals also have to be consistent with the relativistic rigidity condition in (\ref{11d}) for the instantaneous rest frame at the beginning of each transition interval.  With this condition ($|\Delta u|/c \ll 1$), the right-hand side of (\ref{33}) simplifies to
\bea{34}
\f{W_{TI}}{mc^2} &=&\frac{1}{mc^2}\int\limits_{\mbox{\scriptsize TIs}} \left[m\tau_e \gamma^6 \dot{u}^2-\left(f_{a1}+f_{a2}\right)u\right] dt \nonumber\\  
&=& \f{\tau_e}{c}\f{\Delta u_1^{}}{c} \dot{u}(t_1^+)-\f{\gamma(t_2) u(t_2)}{c} \bigg[\f{\tau_e}{c} \gamma^3(t_2) \dot{u}(t_2)\nonumber\\ &&+ \; \gamma^2(t_2) \f{\Delta u(t_2)}{c}\bigg]+\, O\left[(\Delta u/c)^2\right] 
\eea
where $\Delta u(t_2)$ denotes the jump in velocity of the center of the charged sphere across the second transition interval as measured in the laboratory inertial reference frame, which is the rest frame of the sphere before the external force is applied, that is, the rest frame of the sphere at the beginning of the first transition interval ($t=t_1=0$). The $\Delta u_1^{}$ denotes the jump in velocity across the first transition interval as measured in the rest frame of the sphere at the beginning of the first transition interval (the laboratory frame).   Relativistic transformations of acceleration and velocity show that $\gamma^3(t_2) \dot{u}(t_2)$ and $\gamma^2(t_2) \Delta u(t_2)$ are equal, respectively, to $\dot{u}_2^{}$ and approximately to $\Delta u_2^{}$ for $|\Delta u_2|/c \ll 1$, where $\dot{u}_2^{}$ is the acceleration of the sphere in the rest frame at the beginning of the second transition interval (moving with velocity $u(t_2)$ with respect to the laboratory frame), and $\Delta u_2^{}$ is the jump in velocity across the second transition interval as measured in this rest frame at the beginning of the second transition interval.  Thus, (\ref{34}) can be rewritten as
\bea{35}
\f{W_{TI}}{mc^2}&=& \frac{1}{mc^2}\int\limits_{\mbox{\scriptsize TIs}} \left[m\tau_e \gamma^6 \dot{u}^2-\left(f_{a1}+f_{a2}\right)u\right] dt \nonumber\\ 
&=& \f{\tau_e}{c}\f{\Delta u_1^{}}{c} \dot{u}_1^{+}-\f{\gamma(t_2) u(t_2)}{c} \left[\f{\tau_e\dot{u}_2^{}}{c}   +  \f{\Delta u_2^{}}{c}\right]\nonumber\\ && \hspace{30mm}+ \, O\left[(\Delta u/c)^2\right] 
\eea
where $\dot{u}_1^{+} =\dot{u}(t_1^+)$.
\par
If we assume that, like the change in velocity caused by the pre-acceleration and pre-deceleration in the original equation of motion, the change in velocity $\Delta u$ across a transition interval will have the same sign as the change in acceleration ($\dot{u}^{+} -\dot{u}$) across the transition interval,\footnote{One can see that this is the only consistent way to choose $\Delta u$ by noting that the acceleration across any transition interval can be written as the sum of an analytic acceleration formed by analytically continuing the acceleration from its values  before the transition interval (for which there is no change in velocity) and an acceleration that jumps from a value of zero to ($\dot{u}^{+} -\dot{u}$) at the beginning of the transition interval (for which there is a change in velocity in the direction of ($\dot{u}^{+} -\dot{u}$) across the transition interval); see (\ref{24})--(\ref{27}) and \cite[pp. 85--89]{YB}.}   then $\Delta u_1^{}$ will have the same sign as $\dot{u}_1^{+}$ and $\Delta u_2^{}$ will have the opposite sign as $\dot{u}_2^{}$.  Consequently, $\Delta u_1^{} \dot{u}_1^{+} \ge 0$ and under the assumption that
\be{36}
\f{\tau_e|\dot{u}_2^{}|}{c} \ll 1
\ee
for times outside the transition intervals,
one can choose a value of $|\Delta u_2^{}|/c \ll 1$ that is slightly less than or slightly greater than $\tau_e|\dot{u}_2^{}|/c$ to ensure that the energy radiated across the transition intervals is equal to or greater than zero regardless of the sign of $u(t_2)$ or $\dot{u}_2^{}$. With $\tau_e|\dot{u}_1^{+}|/c$ on the order of $|\Delta u_1^{}|/c  \ll 1$, the first term after the second equal sign in (\ref{35}) becomes $O\left[(\Delta u/c)^2\right]$.
\par
If the external force possesses nonanalytic points in time, in addition to when it is first applied and terminated, such that there are a total of $N$ nonanalytic points in time, the total energy radiated across all the transition intervals is given by
\bea{381}
\f{W_{TI}}{mc^2}&=& \frac{1}{mc^2}\int\limits_{\mbox{\scriptsize TIs}} \left[m\tau_e \gamma^6 \dot{u}^2-u\sum_{n=1}^N f_{an}\right] dt \nonumber\\
&=&\f{1}{c^2}\sum_{n=1}^N \gamma(t_n) u(t_n) \left[\tau_e (\dot{u}_n^+ - \dot{u}_n) -\Delta u_n\right]\nonumber\\ &&\hspace{30mm} +\; O\left[(\Delta u/c)^2\right]
\eea
with $\tau_e |\dot{u}^{}_n|/c$ and $\tau_e |\dot{u}^{+}_n|/c$ on the order of  $|\Delta u_n^{}|/c  \ll 1$,
where $\dot{u}_n$ and $\dot{u}_n^+$ are the accelerations of the center of the charged sphere at the beginning and end of the $n$th transition interval as measured in the rest frame of the beginning of the transition interval, and $\Delta u_n$ is the jump in velocity of the center of the charged sphere as measured in this same rest frame.  The $\Delta u_n$ can be chosen to have the same sign as $(\dot{u}_n^+ - \dot{u}_n)$ and slightly less than or greater than $\tau_e (\dot{u}_n^+ - \dot{u}_n)$ (depending on the signs of  $(\dot{u}_n^+ - \dot{u}_n)$ and $u(t_n)$) to keep each term in the second summation of (\ref{381}) equal to or greater than zero, under the rest-frame conditions
\be{382}
\f{\tau_e |\dot{u}_n|}{c} \ll1\,,\;\;\;\;\;n=1,2, \cdots N
\ee
or, more generally, because the $t_n$ may take on any values
\be{383}
\f{\tau_e |\dot{u}(t)|}{c} \ll 1
\ee
in the instantaneous rest frames.
\par
Similarly, it can be shown by integrating (\ref{21}) that under the inequality in (\ref{383}), the total momentum ($G_{TI}$) radiated across all the transition intervals is given by  
\bea{384}
\f{G_{TI}}{mc}&=& \frac{1}{mc}\int\limits_{\mbox{\scriptsize TIs}} \left[m\tau_e \gamma^6 \dot{u}^2 u/c^2 -\sum_{n=1}^N f_{an}\right] dt \nonumber\\
&=&\f{1}{c}\sum_{n=1}^N \gamma(t_n) \left[\tau_e (\dot{u}_n^+ - \dot{u}_n) -\Delta u_n\right]\nonumber\\ &&\hspace{25mm} +\; O\left[(\Delta u/c)^2\right].
\eea
The ratio of each of the terms of the radiated energy and radiated momentum in (\ref{381}) and (\ref{384}) is equal to $u(t_n)$, the velocity of the center of the charged sphere at the beginning of each transition interval --- a result that is physically reasonable for $|\Delta u_n|/c \ll 1$.
\par
Inserting $m_{\rm es}$ from (\ref{2}) into (\ref{18}) shows that as the radius $a$ of the sphere becomes small, $m \approx m_{\rm es} = e^2/(8\pi\eps_0^{}ac^2)$ {\em if the mass is not renormalized with an increasingly negative $m_{\rm ins}$}, and thus from (\ref{19})  we have that $\tau_e \approx 4a/(3c)$.  Then the inequality in (\ref{383}) becomes identical to the one in (\ref{12a}), the inequality needed to ensure that the $O(a)$ terms are negligible in the equation of motion. {\em Consequently, for an extended charged sphere in which the mass is not renormalized as the charge radius is made small, the equation of motion in (\ref{21}) or more generally (\ref{20}) is a causal equation of motion for which the jumps in velocity across the transition intervals  can be chosen to satisfy the relativistic rigidity requirements in (\ref{11rigid}) and to conserve momentum-energy with a non-negative radiated energy during the transition intervals --- provided the conditions in (\ref{12}) are satisfied to ensure that the $O(a)$ terms are negligible.}
\par
The second summations in (\ref{381}) and (\ref{384}) reveal that the  momentum-energy radiated during the transition intervals is negligible (for $|\Delta u_n|/c \ll 1$) if the jumps in velocity ($\Delta u_n$) across the transition intervals are chosen such that
\begin{subequations}
\label{385}
\be{385a}
\Delta u_n = \tau_e \left(\dot{u}_n^+ -\dot{u}_n\right)\,, \;\;\;\; n=1,2, \cdots N
\ee
in the rest frame of the beginning of each transition interval.  In the laboratory frame (the rest frame of the charged sphere before the external force is applied), (\ref{385a}) becomes
\be{385b}
\Delta u(t_n) = \tau_e \gamma(t_n)\left[\dot{u}(t_n^+) -\dot{u}(t_n)\right]\,, \;\;\; n=1,2, \cdots N.
\ee
\end{subequations}
If one assumes that the charged sphere will move in a way to minimize the energy radiated across each transition interval (that is, reduce it to O$[(\Delta u_n/c)^2]$ for $|\Delta u_n|/c \ll 1$), then the changes in velocity across the transition intervals given in (\ref{385}) are mandatory under the conditions in (\ref{11rigid}) and (\ref{12}) for which the equation of motion is derived.  (Recall that $\dot{u}(t_n)$ and $\dot{u}(t_n^+)$ are determined solely by the externally applied force and not by the transition forces or the jumps in velocity across the transition intervals.)  Despite the elegance of choosing $\Delta u(t_n)$ in \eqref{385b} to make the momentum-energy radiated across each transition interval negligible, such a choice is merely based on the conjecture that the charge moves across each transition interval with a change in momentum-energy that becomes reversible as its radius shrinks to a small enough value.  It doesn't appear to be a necessary choice.
\section{EQUATION OF MOTION OF A POINT CHARGE WITH RENORMALIZED MASS}
We have shown that the solutions to the equation of motion in (\ref{20})--(\ref{21}) of a charged sphere of radius $a$ and  rest mass $m \approx m_{\rm es} = e^2/(8\pi\eps_0^{}ac^2)$ are both causal and consistent with conservation of momentum-energy under the conditions on the velocity and its time derivatives given in (\ref{11rigid}) and (\ref{12}) required to derive the equation of motion.  Although these inequalities in (\ref{11rigid}) and (\ref{12}) are  satisfied for all values of the velocity and its derivatives as $a\to 0$, the value of the resst mass $m = m_{\rm es} +m_{\rm ins}$ becomes infinite because the value of $m_{\rm es}$ becomes infinite as $a\to 0$.  It thus seems natural to follow the suggestion of Dirac \cite{Dirac} and simply renormalize the rest mass of the charged sphere as $a\to 0$ to a value $m$ equal to the measured mass of the resulting point charge.  The result of this renormalization of the mass to a finite value as $a\to 0$ takes the same form as the equation of motion in (\ref{20}), namely, in four-vector notation
\be{37}
\f{F_\mathrm{ext}^i\! +\! f_{a1}^i\! + \!f_{a2}^i}{mc^2} = \frac{d u^i}{d s}
-\tau_e\left(\frac{d ^2 u^i}{d s^2}+ u^i\frac{d u_j}{d s}\frac{d u^j}{d s} 
\right).
\ee
The $O(a)$ terms are now perfectly zero because $a\to 0$ and (\ref{12}) are satisfied for all values of the velocity and its derivatives.  In addition, the relativistic rigidity conditions in (\ref{11rigid}) are satisfied.  However, because of the renormalization of the mass $m$, the value of the time constant $\tau_e$ is no longer given by $4a/(3c)$  as $a\to 0$ but by $e^2/(6\pi\eps_0^{} m c^3)$ in (\ref{19}) with the value of $m$ equal to the renormalized mass.  This means that the rest-frame condition in (\ref{383}) required for the conservation of momentum-energy with a non-negative radiated energy across the transition intervals is no longer equivalent to the condition in (\ref{12a}) but must be written as
\begin{subequations}
\label{38}
\be{38a}
\f{e^2}{6\pi\eps_0^{}mc^4}|\dot{u}| \ll 1
\ee
for times outside the transition intervals or, equivalently
\be{38b}
\f{e^2}{6\pi\eps_0^{} m^2 c^4}|\vF_{\rm ext}| \ll 1
\ee
\end{subequations}
in the instantaneous rest frames, where now $m$ is the renormalized mass.
\par
The inequality in (\ref{38a}) or \eqref{38b} is an extra condition that the mass-renormalized charged sphere  must obey as $a\to 0$ in order for the energy in \eqref{381} radiated across the transition intervals to be greater than zero for all possible values of the velocity at the beginning of each transition interval and all possible values of the accelerations just before and just after each transition interval.  These inequalities in \eqref{38} imply from \eqref{381} that the solutions to the renormalized equation of motion in (\ref{37}) for a point charge do not, in general,  remain both causal and consistent with conservation of momentum-energy if the acceleration outside the transition intervals or, equivalently, the applied external force gets too large. In other words, the {\em mass-renormalized} causal classical equation of motion of a {\em point} charge encounters a high acceleration catastrophe.  (One can confirm that such a failure occurs in the solution (\ref{25})--(\ref{26}) to the causal equation of motion of a point charge with renormalized mass $m$ moving through the uniform electric field of a parallel-plate capacitor.) 
\par
There is some justification, even in classical physics, for renormalizing the mass $m_\mathrm{es} +m_\mathrm{ins}$ to a finite value $m$ as $a\to 0$ and $m_\mathrm{es}=e^2/(8\pi\epsilon_0 a c^2) \to \infty$ to obtain the equation of motion of a point charge.  It was mentioned in Footnote 2 that $m_\mathrm{ins}$ may be negative because it can include gravitational and other attractive formation energies.   Thus, as $a\to 0$ it is conceivable that $m_\mathrm{ins} \to -\infty$ and that $\lim_{a\to \infty}(m_\mathrm{es}+m_\mathrm{ins}) =m$, the measured rest mass.  It is especially noteworthy, therefore, that for the point-charge renormalized causal equation of motion in (\ref{37}),  the restriction in (\ref{38a}) on the magnitude of the acceleration, or in (\ref{38b}) on the magnitude of the externally applied force,  is needed to ensure this equation of motion satisfies conservation of momentum-energy while keeping the value of the energy radiated during the transition intervals equal to or greater than zero.\footnote{It may be helpful to restate the reason for this restriction on the magnitude of the externally applied force as the scale factor ($\tau_e$), which is set by the fixed physical mass ($m$), remaining a nonzero constant rather than approaching zero along with the transition intervals as $a\to 0$.  Still, as $a\to 0$, the electrostatic mass approaches an infinite value and, regardless of the justification, replacing this infinite value by the finite value $m$ that sets the scale factor to a nonzero constant is generally referred to as ``renormalization."  A simple way to understand the difference in behavior between the unrenormalized and renormalized solutions is to first note that the equation of motion in either case has the identical form if the $O(a)$ terms are neglected in the unrenormalized case; compare (22b) with (45).  Then for a given left-hand side, (22b) and (45) are identical except for the fact that $\tau_e =4a/(3c)$ in (22b), the unrenormalized case, and $\tau_e = e^2/(6\pi \eps_0^{} m c^3)$ with $m$ fixed in (45) for the renormalized case.  Thus, in the unrenormalized equation of motion, unlike the renormalized equation of motion, the contribution of the radiation reaction term (the term multiplied by $\tau_e$) in the equation of motion approaches zero as the radius $a$ of the charge approaches zero.}
\par
For an electron in an external electric field $E$, the inequality in (\ref{38b}) is satisfied unless $E \not\ll  6\pi\epsilon_0 m^2 c^4/e^3 = 2.7 \times 10^{20}$ Volts/meter, an enormously high electric field.
  Nonetheless,  an equation of motion of a mass-renormalized point charge that is both causal and conserves momentum-energy while avoiding a negative radiated energy during the transition intervals no matter how large the value of the externally applied force does not result by simply equating the sum of the point-charge radiation reaction force and the externally applied force to the relativistic Newtonian acceleration force (renormalized mass multiplied by the relativistic acceleration) and inserting generalized point-function transition forces at the nonanalytic points in time of the external force to obtain (\ref{37}).   A fully satisfactory classical equation of motion of a point charge does not result from the transition-interval-corrected equation of motion for an extended charged particle (an equation that is consistent with causality and conservation of  momentum-energy) by simply renormalizing the diverging electrostatic mass to a finite value as the radius of the charge is allowed to approach zero.
\par
Ultimately, a fully satisfactory equation of motion of a mass-renormalized point charge may require a unified theory of inertial and electromagnetic forces as well as the introduction of quantum effects.  Renormalization of the mass of the charged sphere as its radius shrinks to zero is an attempt to extract the equation of
motion of the point ``electron" from the classical self electromagnetic forces of an
extended charge distribution.  Such attempts, as Dirac wrote \cite{Dirac},
``bring one up against the problem of the structure of the electron,
which has not yet received any satisfactory solution."  
%
\appendix
\section{REQUIREMENTS OF RELATIVISTIC (BORN) RIGIDITY}
In accordance with the approach used in \cite{Born}--\cite{Eriksen-et-al}, consider the points on the diameter of a relativistically rigid sphere of radius $a$ along the $x_0^{}$ axis of any instantaneous inertial rest frame $K_0$ whose origin lies at the center of the sphere.  Denote the position in any $K_0$ frame of each of the points on the diameter (which can be viewed as a rigid rod)  by  $\xi_0$, so that $-a \le \xi_0 \le +a$ with $\xi_0$ independent of time.  Assume that the diameter of the sphere is moving rectilinearly along the $x$ axis of a laboratory inertial frame $K$.  (The $x$ axis of the $K$ frame is collinear with the  $x_0^{}$ axis of any of the $K_0$ frames.) Let $\xi(t)$ denote the position of any point at time $t$ on the diameter of the sphere in the $K$ frame corresponding to the instantaneous rest-frame point $\xi_0$.  The $K$ frame position $\xi(t)$ is a function of $\xi_0$ and thus can be rewritten more precisely as
\be{A1}
\xi(t) = \rmf(\xi_0,t)\,.
\ee
Thus, the velocity of each point on the diameter of the sphere in the $K$ frame is
\be{A2}
u(\xi_0,t) =\f{\partial \rmf(\xi_0,t)}{\partial t}\,.
\ee
\par
The differential separation distance $d\xi(t)$ between two points in the $K$ frame at the time $t$ in the $K$ frame is given in terms of the differential separation distance $d\xi_0$ between the same two points in the instantaneous rest frame by
\be{A3}
d\xi(t) = \rmf(\xi_0 + d\xi_0,t)-\rmf(\xi_0,t) = \f{\partial \rmf(\xi_0,t)}{\partial \xi_0} d\xi_0\,.
\ee
However, $d\xi(t)$ is also related to $d\xi_0$ in the rest frame through the Lorentz relativistic contraction
\be{A4}
d\xi(t) = \sqrt{1-\f{u^2(\xi_0,t)}{c^2}}\;d\xi_0
\ee
so that from (\ref{A3})
\be{A5}
\f{\partial \rmf(\xi_0,t)}{\partial \xi_0} = \sqrt{1-\f{u^2(\xi_0,t)}{c^2}}
\ee
which implies that
\be{A6}
\f{\partial \rmf(\xi_0,t)}{\partial \xi_0} \ge 0\,.
\ee
Insertion of $u(\xi_0,t)$ from (\ref{A2}) into (\ref{A5}) yields the nonlinear differential equation for $\rmf(\xi_0,t)$
\be{A7}
\left[\f{\partial \rmf(\xi_0,t)}{\partial \xi_0}\right]^2 +\f{1}{c^2}\left[\f{\partial \rmf(\xi_0,t)}{\partial t}\right]^2 =1 \,.
\ee
\par
Assume that the sphere is at rest in the $K$ frame until $t=0$ when it begins to move such that the point $\xi$ on the diameter of the sphere has a relativistic  acceleration given by
\be{A8}
\f{d}{dt}\left[\gamma \f{d\xi}{dt}\right] = \f{\partial}{\partial t}\left[\gamma \f{\partial \rmf(\xi_0,t)}{\partial t}\right] = A(\xi_0) \ge 0
\ee
where $A(\xi_0)$ is independent of time $t$.  (Such a uniform acceleration in the $K$ frame characterizes an arbitrary acceleration of the center of the sphere as the radius $a$ of the sphere approaches zero.)\footnote{Although relativistic rigidity requires, in general, that $A$ be a function of $(\xi_0,t)$, the function $A(\xi_0,t)$ can be expanded in a Taylor series about $t=0$ to give $A(\xi_0,t) = A(\xi_0,0) + O(t)$.  Carrying through the derivation with $O(t)$ included adds $O(t^3)$ to the function $\rmf(\xi_0,t)$ given in (\ref{A9a}) with $A(\xi_0,0)$ replacing $A(\xi_0)$ and $A(0,0)$ replacing $A(0)$ in (\ref{A9b})--(\ref{A16}).  Thus, as $t\to 0$, the results in (\ref{A13})--(\ref{A16}) hold in the instantaneous rest frame with $A(\xi_0,t\to 0)$ replacing $A(\xi_0)$, again leading to (\ref{8}) and (\ref{9}).}
\par
The solution to (\ref{A8}) compatible with (\ref{A7}) is 
\begin{subequations}
\label{A9}
\be{A9a}
\rmf(\xi_0,t) = -\xi_s + \sqrt{(\xi_0 +\xi_s)^2 +c^2 t^2}\;,\;\;\;\; t\ge 0
\ee
with
\be{A9b}
A(\xi_0) = \f{c^2}{\xi_0 +\xi_s}
\ee
\end{subequations}
where $\xi_s$ is a constant independent of $\xi_0$ and $t$.   The inequality given in (\ref{A6}) applied to (\ref{A9a}) shows that
\be{A10}
\xi_0 + \xi_s \ge 0\,,\;\;\;\; -a \le \xi_0 \le +a
\ee
which implies that $\xi_s \ge a$ and confirms that the solution in (\ref{A9a}) is for $A(\xi_0) \ge 0$.  The velocity of a point $\xi(t)$ in the $K$ frame is given by
\be{A11}
u(\xi_0,t) =\f{\partial \rmf(\xi_0,t)}{\partial t} = \frac{c^2 t}{\sqrt{(\xi_0 +\xi_s)^2 +c^2 t^2}} \ge 0\,,\;\;\; t \ge 0
\ee
which confirms that the velocity of every point on the diameter of the sphere is zero at $t=0$.
\par
The acceleration of the center of the sphere ($\xi_0 = 0$) is given from (\ref{A9b}) as
\be{A12}
A(0) = \f{c^2}{\xi_s}
\ee
so that the acceleration in (\ref{A9b}) of any other point on the diameter of the sphere can be written as
\be{A13}
A(\xi_0) = \f{A(0)}{1+ \xi_0 A(0)/c^2}\,.
\ee
The left end of the diameter of the sphere will have acceleration equal to 
\be{A14}
A(-a) = \f{A(0)}{1- a A(0)/c^2}
\ee
and the right end
\be{A15}
A(a) = \f{A(0)}{1+ a A(0)/c^2}\,.
\ee
Therefore, only if the acceleration $A(0)$ of the center of the sphere is much smaller than $c^2/a$ can the acceleration of the entire sphere be accurately described by the acceleration of its center. For example, if the acceleration of the center of the sphere is equal to $c^2/a$, the acceleration of the left end of the sphere will be infinite and the acceleration of the right end will equal one half the center value. The velocity in (\ref{A11}) for $t>0$ will also vary rapidly over the sphere unless $A(0) \ll c^2/a$. For small values of $t$, the velocity of the center of the sphere is much less than $c$ ($\gamma \to 1$ as $t\to 0$) and the restriction on the magnitude of the acceleration also applies to the acceleration in the instantaneous rest frame, as given in (\ref{9}), since $t=0$ in this Appendix can be chosen as the time $t$ in (\ref{8}) and (\ref{9}).
\par
Expanding (\ref{A13}) about the center of the sphere ($\xi_0 =0$) produces
\be{A16}
A(\xi_0) = A(0)\left[1 - \f{\xi_0 A(0)}{c^2} + O\left[(\xi_0 A(0)/c^2)^2\right]\right]
\ee
which agrees with \cite[eq. (A.3)]{Newman&Janis} and leads to (\ref{8a}).
%
%
\begin{acknowledgments}
This work benefitted from discussions with Professor Emeritus F. Rohrlich of Syracuse University, Professor T.T. Wu and Dr. J.M. Myers of Harvard University, and Professor W.E. Baylis of the University of Windsor, Canada, and from the comments and suggestions of an anonymous reviewer.  The research was supported in part through Dr. A. Nachman of the U.S. Air Force Office of Scientific Research (AFOSR).
\end{acknowledgments}
\end{document}